\documentclass[%
 reprint,
superscriptaddress,
 amsmath,amssymb,
 aps,
prb,
]{revtex4-2}
\usepackage{siunitx,upgreek}
\usepackage[textsize=tiny]{todonotes}
\usepackage{soul}
\usepackage{color}
\usepackage{gensymb}
\usepackage{placeins}
\usepackage{graphicx}% Include figure files
\usepackage{dcolumn}% Align table columns on decimal point
\usepackage{bm}% bold math
\usepackage[mathlines]{lineno}% Enable numbering of text and display math
\usepackage{mathtools}
\usepackage{amsmath}
\usepackage{comment}
\def\doublecolumnwidth{17.2cm}

\begin{document}

\preprint{APS/123-QED}

\title{Magnetic ordering in out-of-plane artificial spin systems based on the Archimedean lattices}% 

\author{A.~Pac}
\email{aleksandra.pac@epfl.ch}
\altaffiliation[\\
Present address: ]
{Hybrid Quantum Circuits Laboratory (HQC), Institute of Physics, \'Ecole Polytechnique F\'ed\'erale de Lausanne (EPFL), CH-1015 Lausanne, Switzerland}
\affiliation{Laboratory for Mesoscopic Systems, Department of Materials, ETH Zurich, 8093 Zurich, Switzerland}
\affiliation{PSI Center for Neutron and Muon Sciences, 5232 Villigen PSI, Switzerland}

\author{G.~M.~Macauley}
\email{gavin.macauley@psi.ch}
\altaffiliation[\\
Present address: ]
{Department of Physics, Princeton University, Princeton, NJ 08540 USA}
\affiliation{Laboratory for Mesoscopic Systems, Department of Materials, ETH Zurich, 8093 Zurich, Switzerland}
\affiliation{PSI Center for Neutron and Muon Sciences, 5232 Villigen PSI, Switzerland}

\author{J.~R.~Massey}
\affiliation{Laboratory for Mesoscopic Systems, Department of Materials, ETH Zurich, 8093 Zurich, Switzerland}
\affiliation{PSI Center for Neutron and Muon Sciences, 5232 Villigen PSI, Switzerland}

\author{A.~Kurenkov}
\affiliation{Laboratory for Mesoscopic Systems, Department of Materials, ETH Zurich, 8093 Zurich, Switzerland}
\affiliation{PSI Center for Neutron and Muon Sciences, 5232 Villigen PSI, Switzerland}

\author{F.~Mila}
\affiliation{Institute of Physics, \'Ecole Polytechnique F\'ed\'erale de Lausanne (EPFL), CH-1015 Lausanne, Switzerland}

\author{P.~M.~Derlet}
\affiliation{Laboratory for Mesoscopic Systems, Department of Materials, ETH Zurich, 8093 Zurich, Switzerland}
\affiliation{PSI Center for Scientific Computing, Theory and Data, 5232 Villigen PSI, Switzerland}

\author{L.~J.~Heyderman}
\email{laura.heyderman@psi.ch}
\affiliation{Laboratory for Mesoscopic Systems, Department of Materials, ETH Zurich, 8093 Zurich, Switzerland}
\affiliation{PSI Center for Neutron and Muon Sciences, 5232 Villigen PSI, Switzerland}

\date{\today}% It is always \today, today,

\begin{abstract}
Artificial spin systems, sometimes referred to as artificial spin ices, are arrays of coupled nanoscale magnets that order according to the lattice geometry, nanomagnet shape and magnetic anisotropy.
Here we characterize a family of artificial spin systems that are formed by placing arrays of out-of-plane nanomagnets on the vertices of the Archimedean lattices. 
On demagnetizing these nanomagnet arrays using a magnetic field protocol and subsequently imaging the magnetic configuration using magnetic force microscopy, we observe different types of magnetic order. 
We compare our experimental results with those predicted by Monte Carlo simulations to assign an effective temperature to each lattice.
We find that, for all of the lattices, the assigned effective temperature is above the transition temperature.
This reflects the difficulty of obtaining system-spanning order in lattices with out-of-plane nanomagnets.
We consider to what extent further-neighbor interactions affect the phase diagram and spin-spin correlations in each lattice, illustrating our results with four example lattices.
We can divide the lattices into three main categories: bipartite lattices that admit a perfect antiferromagnetic ground state, frustrated lattices where ordering proceeds via a single step, and  frustrated lattices with two-step-ordering.
Our work highlights the diversity of magnetic ordering that can be hosted in two-dimensional artificial spin systems with out-of-plane nanomagnets, and demonstrates the importance of including long-range interactions to explain the magnetic ordering. 
Such insights will be important for incorporating 
artificial spin systems into novel computing applications.
\end{abstract}

\maketitle

%\tableofcontents

\section{\label{sec:introduction}Introduction}

Artificial spin systems are arrays of dipolar-coupled nanomagnets~\cite{wang, sandra} that exhibit interesting phenomena including spin liquid~\cite{2016Canals} and spin glass physics~\cite{2019Saccone, 2022Saccone}, vertex frustration~\cite{2016Gilbert}, and emergent magnetic monopoles~\cite{2010Ladak, 2011Mengotti}.
They were initially envisaged as a two-dimensional~(2D) analogue to frustrated rare earth spin-ice compounds~\cite{harris}, themselves a magnetic counterpart to water ice, in which incommensurate bonding distances between hydrogen ions and oxygen centres give rise to a residual entropy~\cite{pauling}.
In this framework, artificial spin systems are often referred to as artificial spin ices because they demonstrate similar local constraints, short-range correlations and macroscopic ground state degeneracy. 
However, not all artificial spin systems display ice-like physics, but rather manifest a wide variety of emergent behaviours that depend on the lattice geometry, nanomagnet shape and magnetic anisotropy.
Typically, artificial spin systems are $2$D~lattices of single-domain nanomagnets, whose magnetic configurations can be directly accessed in real space, with magnetic force microscopy (MFM)~\cite{wang}, magneto-optical Kerr effect (MOKE) microscopy~\cite{MOKE}, and synchrotron x-ray or electron microscopy~\cite{2009Mengotti,2011Mengotti,LTEM1,LTEM2}, and also in reciprocal space, with x-ray scattering~\cite{2019Sendetskyi}, for example.
Analysis of these magnetic configurations is often simplified by assuming that each single-domain nanomagnet can be represented as a macrospin, with a direction set by the direction of the net moment of the nanomagnet.

Beyond their fascinating physics, artificial spin systems have also attracted interest as a way to perform logic operations~\cite{logic, 2018Arava}, and for integration into unconventional computing schemes such as neuromorphic computing~\cite{heydermancomputing,jack,Zhaochu, 2024Kurenkov}. 
For such applications, arrays of nanomagnets with perpendicular magnetic anisotropy~\cite{2009Mengotti, 2012Zhang, 2014Chioar, 2017Fraleigh, 2020Kempinger_stochastic} are of particular interest because the magnetic state of the individual nanomagnets can be accessed via electrical means using the anomalous Hall effect~\cite{hall,ahe,sala,sala_sot}.
This allows the direct reading or writing of the state of each nanomagnet using current, thus making it possible to interface artificial spin systems with existing CMOS architectures. 
However, before such out-of-plane artificial spin systems can be exploited in applications, it is important to have a greater understanding of the effect of the lattice geometry and the long-range dipolar coupling on the resulting physics.
The fine balance between these two factors sets the hierarchy of intermagnet interactions, or weights~\cite{2019Chakraverty}, which is interesting for designing neuromorphic networks~\cite{hanu,2016Grollier,2021Yang,2020Song}.

Here we present a complete experimental and computational classification of the behaviour of out-of-plane artificial spin systems based on the Archimedean lattices~\cite{1987Grunbaum}, which are shown in Fig.~\ref{fig:SEM}. 
As can be seen in this figure and also in the Supplemental Material, Section~1~\cite{Supplementary}, the Archimedean lattices are $11$ uniform $2$D~lattices made from regular polygons, such that every vertex is topologically equivalent.
In all of the lattices that we fabricated, the base nanomagnet has the same diameter and is made of the same magnetic material. 
A given set of $11$~Archimedean lattices with the same nanomagnet separation thus only differ in the number and positions of neighboring macrospins.
This allows us to compare the emergent magnetic order resulting from the dipolar interactions across the whole Archimedean family by only changing the lattice geometry.
For each lattice, we determine the magnetic state following a magnetic field demagnetization, and compare the observed spin-spin correlations with those predicted by simulated anneals using Monte Carlo simulations.
This allows us to assign an effective temperature to each lattice, quantifying the extent to which the low-temperature spin arrangements can be accessed experimentally. 
In particular, by comparing the effective temperature of the experimental systems to the critical temperature of each lattice determined with Monte Carlo simulations, we can ascertain how close each artificial spin system gets to its phase transition.
For the most part, we find that these lattices reach an effective temperature slightly above their critical temperature.
Of the $11$~lattices, four are bipartite: the Square, Honeycomb, Square Hexagonal Dodecagonal (SHD) and CaVO lattices.
This means they admit a perfectly-ordered antiferromagnetic ground state.
The remaining seven exhibit varying degrees of frustration, which we quantify through their entropy as determined with Monte Carlo simulations based on single spin flips.
Two of these seven lattices, the Triangular and Kagome lattices, are highly frustrated and their ordering proceeds in two steps.

\begin{figure*}[!htbp] 
    \centering
    \includegraphics[width=\doublecolumnwidth, keepaspectratio]{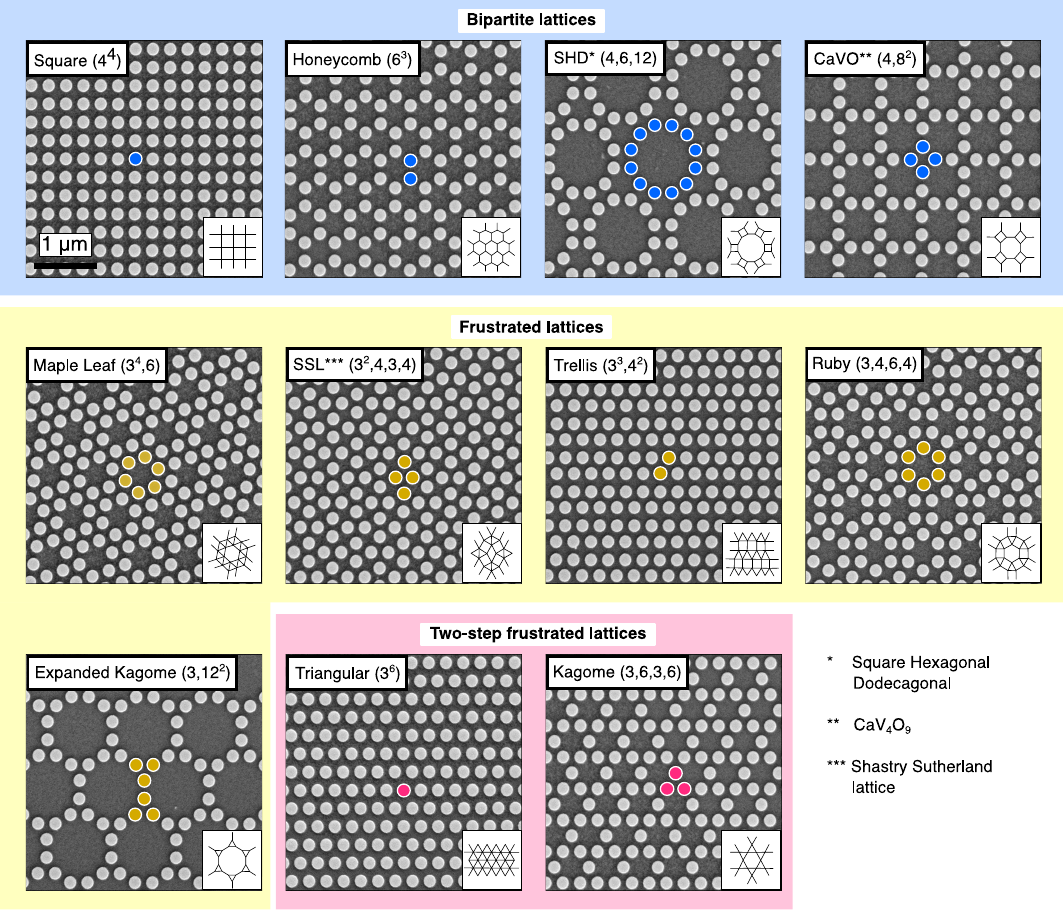}
    \caption{\label{fig:SEM} \textbf{SEM images of artificial spin systems with out-of-plane nanomagnets located on the vertices of the $11$~Archimedean lattices.}
    The name we give to each lattice, as well as the notation of Gr{\"u}nbaum and Shephard~\cite{1987Grunbaum}, which describes the shapes meeting at each lattice point, are given in the top left corner of each panel.
    The building block for all lattices is a Co/Pt nanomagnet with diameter $d = 200$~nm.
   Because of the strong perpendicular magnetic anisotropy, each nanomagnet behaves as a single macrospin with a magnetic moment either pointing out of or into the plane, which we refer to as up or down, respectively.
   While we have created arrays of nanomagnets with lattice parameters going from $220$~nm to $300$~nm, we show here arrays with lattice constant $a = 300$~nm, where the separation between adjacent nanomagnets $s = 100$~nm, because the largest lattice constant allows for better discrimination of each pattern.
    For each lattice, the nanomagnets highlighted in color indicate those in a single unit cell. 
    Each panel contains a drawing of the underlying lattice structure in the bottom right corner.  
    SEM images of the $a = 220$~nm arrays, which are the most densely packed arrays we fabricated with a separation between adjacent nanomagnets of $20$~nm, can be found in the Supplemental Material, Section~1~\cite{Supplementary}.
    In the main text, we present experimental results obtained with the $a = 225$~nm arrays (i.e, with separation $s = 25$~nm).
    These are the arrays of nanomagnets with the strongest dipolar coupling, where we are confident that all of the magnetic material between adjacent nanomagnets has been removed by ion milling.
    Although the nanomagnet arrays with $a = 220$~nm should, in principle, exhibit even an stronger dipolar coupling, the incomplete removal of magnetic material results in some nanomagnets being exchange-coupled with their neighbors. }
\end{figure*}

Because of their strong geometric frustration, the Triangular and Kagome lattices are important model systems in the field of highly frustrated magnetism.
The Triangular lattice with first-nearest-neighbor antiferromagnetic interactions is the archetypal frustrated lattice, first introduced by Wannier~\cite{1950Wannier}.
This lattice was predicted to have a massively-degenerate ground state, with a finite residual entropy $S_{\mathrm{1nn}}^{\mathrm{Tri}} = 0.323~k_B$~per~spin.
In this context, the residual entropy $S_0$ is related to the logarithm of the number of ground states $\Omega$ through $S_0 = (k_B/N) \ln (\Omega)$.
More recently, Smerald~{et al.}~\cite{2018Smerald} demonstrated that, in a Triangular lattice with dipolar interactions, the system crosses over from a weakly-coupled paramagnetic phase at high temperature to a spin-liquid phase as the temperature is decreased. 
Upon further cooling, the system then undergoes a first-order transition into a striped phase.
The ground state for the Triangular lattice with dipolar interactions is six-fold degenerate, which arises from the three-fold degeneracy in terms of the possible stripe directions combined with the two-fold Ising degeneracy associated with each of the spins~\cite{2018Smerald}.
In the spin-liquid phase, almost all of the three-nanomagnet triangular plaquettes have a spin configuration that is either ``two-up/one-down'' or ``two-down/one-up''.
However, because there are a large number of configurations that allow for such a local arrangement of spins, long-range ordering is hindered.
Just above the transition to the striped phase, the entropy of the Triangular lattice with dipolar interactions is approximately $0.22~k_B$~per spin, which is lower than the residual entropy obtained from the first-nearest-neighbor model of $S_{\mathrm{1nn}}^{\mathrm{Tri}} = 0.323~k_B$~per~spin~\cite{1950Wannier}.
In the ground state, the Triangular lattice with dipolar interactions possesses no residual entropy.

Similarly, the thermodynamics of the Kagome lattice has been investigated~\cite{1951Syozi, 1953Kano, 2018Hamp}.
On this lattice, a first-nearest-neighbor antiferromagnetic model admits the existence of only short-range spin-spin correlations at low temperatures based on the formation of low-energy plaquettes~\cite{1951Syozi} as was seen for the Triangular lattice.
The residual entropy for the Kagome lattice with only first-nearest-neighbor antiferromagnetic interactions is $S_{\mathrm{1nn}}^{\mathrm{Kag}} =0.501~k_B$~per spin~\cite{1953Kano}.
When further-neighbor dipolar interactions are included, Monte Carlo simulations revealed a two-stage ordering process in the Kagome lattice, leading to a ground state consisting of alternating stripes~\cite{2018Hamp}. 
  
We have characterized nanofabricated artificial spin systems based on all $11$ Archimedean lattices with MFM and using full-dipolar Monte Carlo simulations. 
Our comprehensive work offers a foundation for future research, particularly since we realize several lattices experimentally for the first time.
By fabricating all of the Archimedean lattices together on one substrate, we have acquired a complete data set with minimal differences in the intrinsic properties of the constituent nanomagnets.
This allows us to compare the effect of different hierarchies of interactions across the family of lattices.

\section{Artificial spin systems based on the Archimedean lattices: background and experimental system}

Archimedean lattices are uniform tilings of the plane in which all vertices are alike, which means that each vertex is surrounded by the same number and type of polygons.
Kepler first proved that there were only $11$~such lattices, and named them in homage to the Archimedean solids~\cite{1987Grunbaum}.
Eight of the Archimedean lattices feature two or or more types of polygon, while the standard uniform tilings (Square, Triangular and Hexagonal) feature only one.
Crystal planes in many real materials, such as alloys, resemble Archimedean lattices~\cite{1980Okeefe}.
In addition, these tilings also appear in fields as diverse as the template-directed assembly of eutectic structures, where diffusion of the fluid imposes symmetry constraints on the final structure~\cite{2020Kulkarni}, in polymer~\cite{2005Takano} and other molecular complexes~\cite{2020Voigt}, and in photonic crystals, where the Archimedean nature of the lattice can promote the formation of a photonic band gap~\cite{2001David, 2007Ueda}.
In addition, the Archimedean lattices can be found in percolation problems, which serve as models for the movement of fluid through a medium and for which the site percolation threshold for many of the lattices remains unsolved~\cite{1999Sudding}. 

In terms of magnetic systems, different Hamiltonians on individual Archimedean lattices have been extensively investigated, with perhaps the most famous being Onsager's solution of the $2$D~Ising model on the Square lattice~\cite{1944Onsager}.
The prototypical frustrated magnetic system---a Triangular lattice with first-nearest-neighbor antiferromagnetic exchange---is another famous example~\cite{1950Wannier}.
Complete catalogues of the magnetic behaviour of all Archimedean lattices are somewhat rarer, although Ref.~\cite{2010Codello}, which lists exact Curie temperatures for the case of a ferromagnetic first-nearest-neighbor interaction, is a notable exception. 
For the case of a Hamiltonian with \emph{antiferromagnetic} first-nearest-neighbor interactions between Ising spins, Yu presented a comprehensive study of the different Archimedean lattices with Monte Carlo simulations and classified their ground states, finding that four of the lattices were unfrustrated while the remaining seven were frustrated to different degrees~\cite{2015yu}.

Taking inspiration from the works highlighted above, we explored the Archimedean lattices experimentally with artificial spin systems comprising arrays of \emph{dipolar-coupled} nanomagnets where every nanomagnet is coupled to every other through the long-range dipolar interaction. 
To fabricate our systems, we patterned a $\mathrm{Ta(4)/Pt(3)/[Co(1.1)/Pt(0.2)]_{2}/Co(1.1)/Ru(2)}$ multilayer stack (numbers in brackets are thicknesses in nm) into arrays of circular nanomagnets placed on the vertices of  the $11$~Archimedean lattices, as shown in Fig.~\ref{fig:SEM}. 
In the Supplemental Material, Section~2~\cite{Supplementary}, we present magnetometry measurements of this multilayer stack. 
These confirm that the magnetisation of the film lies out-of-plane at remanence, with a coercive field of approximately $25.5$~mT.
The nanomagnets in our arrays are circular and have a diameter $d = 200$~nm.
The materials and layer thicknesses in the stack, as well as the diameter of the nanomagnets, have been carefully chosen so that the nanomagnets are single domain with a strong perpendicular magnetic anisotropy.
There is an interfacial Dzyaloshinskii–Moriya interaction at the boundary between the platinum and the cobalt layers, which may contribute to this perpendicular magnetic anisotropy.
However, the exact origin of the perpendicular magnetic anisotropy depends on the materials and the layer thicknesses in the multilayer stack, and can arise from both the broken symmetry at the interfaces and the interfacial Dzyaloshinskii–Moriya interaction~\cite{2011Bandiera}.
By sputtering multiple layers of cobalt, we increase the total volume of magnetic material, leading to a stronger coupling between nanomagnets.
Since each nanomagnet is single domain, it can be represented by an Ising macrospin, with a magnetic moment that points in one of the two out-of-plane directions, up or down.

In such out-of-plane artificial spin systems, the nanomagnets in each array are magnetostatically-coupled so that the interaction between any two neighboring nanomagnets favors their antiparallel alignment.
The total thickness of magnetic material in the stack is $3.3$~nm, from which we estimate the magnetic moment of a single nanomagnet to be $1.5\times 10^7\,\mu_B$.
This sets a lower limit for the interaction strength because it neglects any possible polarisation of the Pt layers.
In addition, the strength of the interaction only depends on the distance between the macrospins.
For the experimental work presented here, we focus on systems with a lattice constant $a = 225$~nm, giving a separation between first-nearest-neighbor nanomagnets of $25$~nm.
These were the most strongly coupled arrays that we fabricated, with the magnetic material between adjacent nanomagnets completely removed by ion milling.
While we also patterned even closer arrays with $a = 220$~nm and $s = 20$~nm (images of which are shown in the Supplemental Material, Section~1~\cite{Supplementary}), some magnetic material was present between adjacent nanomagnets in these arrays.
This remaining material introduced an effective ferromagnetic exchange coupling between adjacent nanomagnets. 
This was evident in the MFM images because the moments of adjacent nanomagnets tended to align in the same direction, leading to an increase in the first-nearest-neigbor spin-spin correlation compared with the $a = 225$~nm arrays (Supplemental Material, Section~4~\cite{Supplementary}).

Based on micromagnetic simulations performed using MuMax3~\cite{2014Vansteenkiste} (detailed in the Supplemental Material, Section~2~\cite{Supplementary}), we estimate that the interaction strength between first-nearest-neighbor nanomagnets, separated by $25$~nm, is  $\approx2.73 \times 10^{-19}$~J.
In addition, we also show how this interaction strength changes as function of lattice constant $225~\mathrm{nm} \leq a \leq 300~\mathrm{nm}$.
The interaction strength falls to $\approx0.78 \times 10^{-19}$~J for nanomagnet arrays with $a = 300$~nm, where the nanomagnets are separated by $100$~nm.
In the Supplemental Material, Section~2~\cite{Supplementary}, we compare the first-nearest-neighbor spin-spin correlations of the nanomagnet arrays presented in the main text, where each cobalt layer has a thickness of $1.1$~nm, to the spin-spin correlations of nanomagnet arrays fabricated from a stack where each cobalt layer has a thickness of $0.9$~nm.
The same field demagnetization protocol is applied to both samples.
Since the interaction strength between nanomagnets scales approximately with the square of their saturation magnetization, the nanomagnets milled from a multilayer stack with $0.9$~nm-thick cobalt layers exhibit weaker coupling to their neighbors.
As a consequence, the magnitudes of the first-nearest-neighbor spin-spin correlations are smaller for this stack, and the nanomagnets in these arrays freeze at a higher effective temperature.
All of the data we present in the main text is for artificial spin systems comprising nanomagnets patterned from the stack with $1.1$~nm-thick cobalt layers and with $a = 225$~nm.

Each array contains $1500$ to $2500$ nanomagnets, and spans an area of $10\times10~\upmu$m$^2$.
Each array was captured in a single MFM scan to determine the magnetic configuration.
The exact number of nanomagnets in an array depends on the geometry of each lattice and thus their areal density.
All nanomagnet lattices are fabricated on the same substrate, with five nominally identical copies of a given lattice with the same lattice constant.
Further details regarding the fabrication are contained in Appendix~A. 

Regarding the naming of our lattices, which is given in Fig.~\ref{fig:SEM}, we tend to adopt the standard name where it is well-accepted, e.g. Triangular, or where a spin system on that lattice has been studied already, as in the case of the Shastry-Sutherland lattice (SSL)~\cite{1981Shastry}, or where it mimics a certain crystal structure, as for the CaVO lattice.
Otherwise, we use the nomenclature of Ref.~\cite{2015yu}, with two exceptions: ``Expanded Kagome'' instead of ``Star'', and ``Ruby'' instead of ``Bounce".
It should also be pointed out that other names are in use, such as ``Briarwood'' or ``Bathroom Tiling'' for the CaVO lattice~\cite{1999Sudding}.
In addition, there exists a general notation after Gr{\"u}nbaum and Shephard, which classifies each Archimedean lattice based on the polygons that surround a vertex~\cite{1987Grunbaum}.
For example, the Trellis lattice is then denoted by $(3^3, 4^2)$ since \emph{three} triangles followed by \emph{two} squares are encountered going clockwise around each vertex.
This notation is also given in Fig.~\ref{fig:SEM}.

To make our results easier to classify, we have sorted the $11$~Archimedean lattices into three groups, as indicated by the blue, yellow and red backgrounds in Fig.~\ref{fig:SEM}:
\begin{enumerate}
\item Bipartite lattices, where the lattice points can be divided into two sublattices such that, for an arbitrary lattice point, its first-nearest-neighbors are all in the other sublattice. 
For our systems, in which the dipolar interaction favors antiparallel alignment between any pair of nanomagnets, this means that this bipartite system can  accommodate a N{\'e}el-like ground state. This group contains the Square, Honeycomb, SHD and CaVO lattices.
\item Frustrated lattices, where a perfect antiferromagnetic ground state cannot be accommodated. This group contains the SSL, Trellis, Ruby, Maple Leaf and Expanded Kagome lattices. 
These lattices are not bipartite because their vertices involve at least one odd-sided polygon, which disrupts any alternating sequence of up/down states as one moves in a loop around the polygon.
\item Two-step frustrated lattices, which are highly frustrated and where, as we will show, the magnetic ordering occurs in two stages. The remaining two lattices, Triangular and Kagome, belong to the two-step frustrated lattices group. One can consider the two-step frustrated lattices as a subset of the generally frustrated ones.
However, we keep them in a separate group because they display fundamentally different thermodynamic behavior in the full dipolar model.
\end{enumerate}

To justify the classification of these lattices into three groups, we now turn to Monte Carlo simulations to elucidate their phase diagrams. 
In particular, by examining the number of peaks that appear in the specific heat capacity as a function of temperature, along with the change in entropy, we provide evidence for our choice of groups.

\section{\label{sec:phase_diagrams}Determining the phase Diagrams of the Archimedean lattices with Monte Carlo simulations}

To capture the thermal behavior of the Archimedean lattices, Monte Carlo simulations were performed using the Metropolis-Hastings algorithm~\cite{1949Metropolis,1953Metropolis} with single spin flip dynamics. 
Each nanomagnet was approximated as an Ising point dipole that can take one of the two possible out-of-plane directions. 
The full details of the Monte Carlo simulations can be found in Appendix~B.

The two limiting cases that we consider are (i) the full-dipolar model and (ii) the first-nearest-neighbor model. 
The first case corresponds to the situation in which every nanomagnet is coupled to every other in the array, irrespective of their separation.
The second case corresponds to a severe truncation of the Hamiltonian, such that each spin is only coupled to its immediate neighbors.
Determining the phase diagrams and the spin-spin correlations as a function of temperature using these two models, and comparing the resulting data, allows us to determine the effect of long-range interactions.
As we will show, for some of the lattices, it is important to include all of the interactions beyond the first-nearest neighbors to be able to obtain a good agreement with the correlations we observe in our experimental samples. 

\begin{figure*}
    \centering
    \includegraphics[width=\doublecolumnwidth, keepaspectratio]{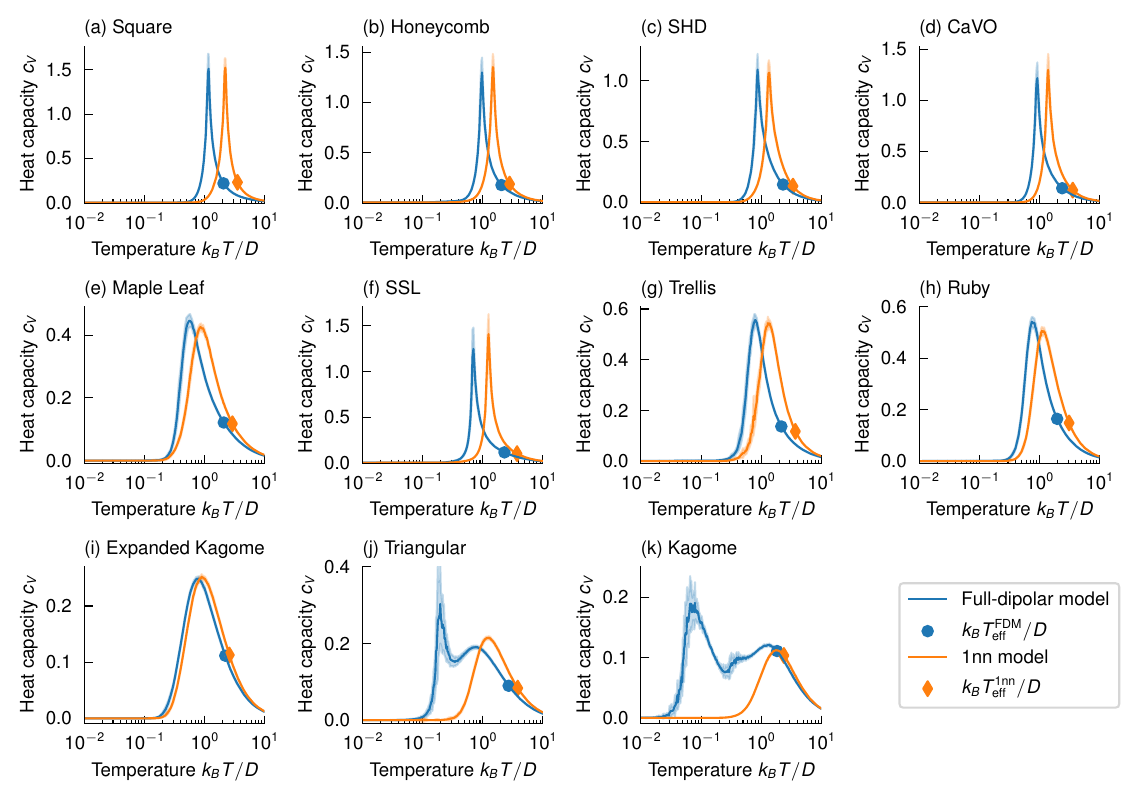}
    \caption{\label{fig:PhaseDiagram} \textbf{Phase diagrams of the $\textbf{11}$~Archimedean lattices determined with Monte Carlo simulations.}
    Heat capacities as a function of temperature for the bipartite lattices [panels (a)-(d)], frustrated lattices [panels (e)-(i)], and two-step frustrated lattices [panels (j) and (k)].
    In each panel, the blue and orange curves correspond to the full-dipolar model and the first-nearest-neighbor model, respectively.
    Peaks in the heat capacity represent phase transitions or crossovers between different types of magnetic order. 
    On including all interactions (i.e, comparing the curves for the full-dipolar model with the curves for the first-nearest-neighbor model), the location of the peaks shifts to lower temperatures.
    For the two-step frustrated lattices, a low-temperature second peak appears on an energy scale well below that of the first-nearest-neighbor interaction in the system, $J_{1\mathrm{nn}} = 1$.
    In general, the temperature at which a crossover or phase transitions occurs is set by the energy of the dominant interaction that causes it, i.e. $T_C \approx J$.  
    This means that the lower-temperature peaks are controlled by spin-spin correlations between further-neighbor pairs, which are weaker in strength.
    For each panel, the blue circle and the orange diamond indicate the average effective temperature of each experimental array when calculated according to the spin-spin correlations of the full-dipolar model and the first-nearest-neighbor model, respectively.
    For space reasons, the units of $c_V$, ($k_B$ spin$^{-1}$),  are omitted from the axis labels in each panel.}
\end{figure*}

For the phase diagrams, we calculate the specific heat capacity $c_V$ as a function of temperature.
Details on these calculations are given in Appendix~B.
Peaks in the heat capacity correspond to either crossovers or phase transitions~\cite{1995Chaikin}, and are located at the corresponding crossover or critical temperature, respectively. 
The heat capacities versus temperature are shown for each lattice in Fig.~\ref{fig:PhaseDiagram} for the full-dipolar model (blue curves) and first-nearest-neighbor model (orange curves). 
In general, the inclusion of all interactions has the effect of shifting the critical temperatures to lower values with the peaks in the blue curves located to the left of the peaks in the orange curves.

For the bipartite lattices, the phase diagrams all contain a single peak [Fig.~\ref{fig:PhaseDiagram}(a)-(d)], indicating that the ordering proceeds in a single step.
In addition, there is no difference between the ground state for the full-dipolar model and the first-nearest-neighbor model.
In both cases, the ground state is a pattern  of nanomagnets with alternating up and down spins.
Example Monte Carlo configurations at $T = 0$ for both the full-dipolar and first-nearest-neighbor models, confirming the nature of the ground state, are provided in Section 5 of the Supplemental Material~\cite{Supplementary}. 
This section also includes a broader discussion of features in the magnetic structure factor for all of the $11$~Archimedean lattices.

The heat capacities of the frustrated lattices are also all single-peaked [Fig.~\ref{fig:PhaseDiagram}(e)-(i)]. 
The magnetic structure factors of the $T = 0$ magnetic configurations for both the full-dipolar model and the first-nearest-neighbor model look remarkably similar. 
These are shown in the Supplemental Material, Section~5~\cite{Supplementary}.
Since the magnetic structure factor encodes the spin-spin correlations, it thus appears that the correlations in the frustrated lattices are practically the same irrespective of the model chosen.

We make some comments about the real-space magnetic configurations of these lattices in the ground state.
While the triangular plaquettes in the SSL and Trellis lattices are frustrated, long-range order does emerge in these lattices in both the full-dipolar and first-nearest-neighbor models. 
This accords with Yu~\cite{2015yu}, who classifies these lattices as frustrated but with long-range-ordered ground states.
The other three frustrated lattices---the Ruby, Maple Leaf and Expanded Kagome---do not exhibit long-range order in either the first-nearest-neighbor model or the full-dipolar model, with regions of diffuse intensity appearing in their magnetic structure factor in the ground state.
In the context of the first-nearest-neighbor model, Yu~\cite{2015yu} described the ground states of these lattices as similar to a spin ice, since macrospins obey local constraints regarding the number of up and down spins at plaquettes.
For the Ruby, Maple Leaf, and Expanded Kagome lattices, the real-space magnetic configurations and magnetic structure factors in the ground state as obtained from Monte Carlo simulations are nearly identical in both the full-dipolar model and the first-nearest-neighbor model. 
This suggests that Yu’s~\cite{2015yu} classification of these lattices as spin-ice-like in the first-nearest-neighbor model remains valid also in the full-dipolar model.

Finally, the phase diagrams of the two-step frustrated lattices, Triangular and Kagome, are shown in Fig.~\ref{fig:PhaseDiagram}(j) and Fig.~\ref{fig:PhaseDiagram}(k), respectively.
Interestingly, both lattices exhibit a second peak for the full-dipolar model, while the second peak is absent when only first-nearest-neighbor interactions are included.
For both lattices, and for both models, there is a broad peak close to $k_B T/D = 1$, which is on the order of the first-nearest-neighbor interaction $J_{1\mathrm{nn}} = 1$. 
When using the full-dipolar model, the second peak observed for both the Triangular and the Kagome lattices is much sharper and occurs at a significantly lower temperature of $k_B T/D \approx 0.1$.

In general, local maxima in the specific heat capacity are suggestive of ether a phase transition or, at least, a crossover in the system in the thermodynamic limit.
For the bipartite lattices, which attain a long-range ordered ground state in either model, the peaks in their heat capacity most likely reflect a second order Ising phase transition.
This has been numerically confirmed for the Square lattice~\cite{1980Racz}, and probably holds for the other bipartite lattices.
For the two-step frustrated lattices, the single peak that appears when only first-nearest-neighbors are considered is a crossover, as established in References~\cite{1950Wannier, 1951Syozi}.
When dipolar interactions are included, Smerald~\emph{et al.}~\cite{2018Smerald} established for the Triangular lattice that the upper peak is a crossover and the lower peak is a first-order phase transition.
To the best of our knowledge, no similar undertaking has been carried out to determine the nature of the phase transitions or crossovers in the Kagome lattice, although we would speculate that they are similar to those of the Triangular lattice. 
For the frustrated lattices, based on the results in Fig.~\ref{fig:PhaseDiagram}, we cannot explicitly define whether the peaks in the heat capacities are phase transitions or crossovers. 
Future work in this area could include undertaking a finite-size analysis of these five lattices to uncover their scaling behavior and, thus, the type of transitions.
We present the first steps towards this endeavor in the Supplemental Material, Section~3~\cite{Supplementary}, where we display the results for a finite-size scaling of the Expanded Kagome lattice.
These results demonstrate two key points: (1) the position of the peak in the heat capacity shows minimal shift for system sizes beyond $600$ spins; and (2) at this size, the effective residual entropy is nearly equal to its extrapolated value in the thermodynamic limit.
Our experimental nanomagnet arrays always contain at least $1500$ nanomagnets, indicating that the system sizes are sufficiently large to avoid the need for finite-size corrections.
Although beyond the scope of the current work, it would be useful in the future to repeat this finite-size scaling for the remaining ten Archimedean lattices and extend this analysis to determine the critical exponents associated with their phase transitions.
Having established the nature of the phase diagram for each Archimedean lattice, we now compare the correlations in the experimental systems after demagnetization with those obtained with a Monte Carlo simulated anneal. 

\section{\label{sec:demag}Magnetic configurations in experimental systems following demagnetization}
The magnetic configuration in many of these Archimedean artificial spin lattices has not been imaged experimentally, and the exact nature of the low-energy configurations that can be reached is not known.
While a few of the lattices---including the Square, Triangular and Kagome lattices---have been investigated~\cite{2012Zhang, 2014Chioar}, we have characterized the behaviour of \textit{all} eleven lattices experimentally as well as with Monte Carlo simulations.
To probe the low energy states of our Archimedean spin lattices, we subjected them to a demagnetization protocol involving an alternating magnetic field, with a maximum applied field of $1$~T and reducing the field in steps of $0.1$~mT to zero field.
At each step, the field was applied for at least $3$~seconds.
The total demagnetization procedure took around 17 hours to complete.
Such a magnetic field protocol is often used to prepare low-energy magnetic configurations in artificial spin systems \cite{wang,2020Kempinger_stochastic}. One could also consider performing a thermal anneal \cite{2011Morgan,2013Zhang}. However, we avoid applying heat to our arrays of out-of-plane nanomagnets because the interfaces between the layers, and therefore the perpendicular magnetic anisotropy, may be modified on heating.

After demagnetization, the magnetic configuration of each lattice was measured using MFM. 
With MFM, each nanomagnet presents with a uniform black or white contrast, which allows for unambiguous assignment of the direction of the associated macrospin, which points up or down, respectively.
We find that the residual magnetization after demagnetization is low (typically less than $10$\% of the saturation magnetization).
Any residual magnetization is likely due to residual stray magnetic fields present in the experimental setup.
In particular, during the demagnetization protocol, the sample is located between the poles of an electromagnet.
To obtain the required maximum field, the gap between the poles is set to $\approx 2$~cm and the field supplied by the electromagnet is calibrated prior to each protocol. 
Even at zero current, there is a small residual field from the cores of the electromagnet, which may slightly bias the magnetic state of our lattice.
This is likely to be responsible for the small residual magnetization even after the protocol.
To minimize the effects of the experimental setup and obtain more data points used for averaging, the demagnetization protocol was repeated three times.
Each lattice appears five times on the sample and thus the spin-spin correlation values determined from the MFM images represent the average of $15$ ($3 \times 5$) measurements.

\begin{figure*}
    \centering
    \includegraphics[width=0.85\textwidth]{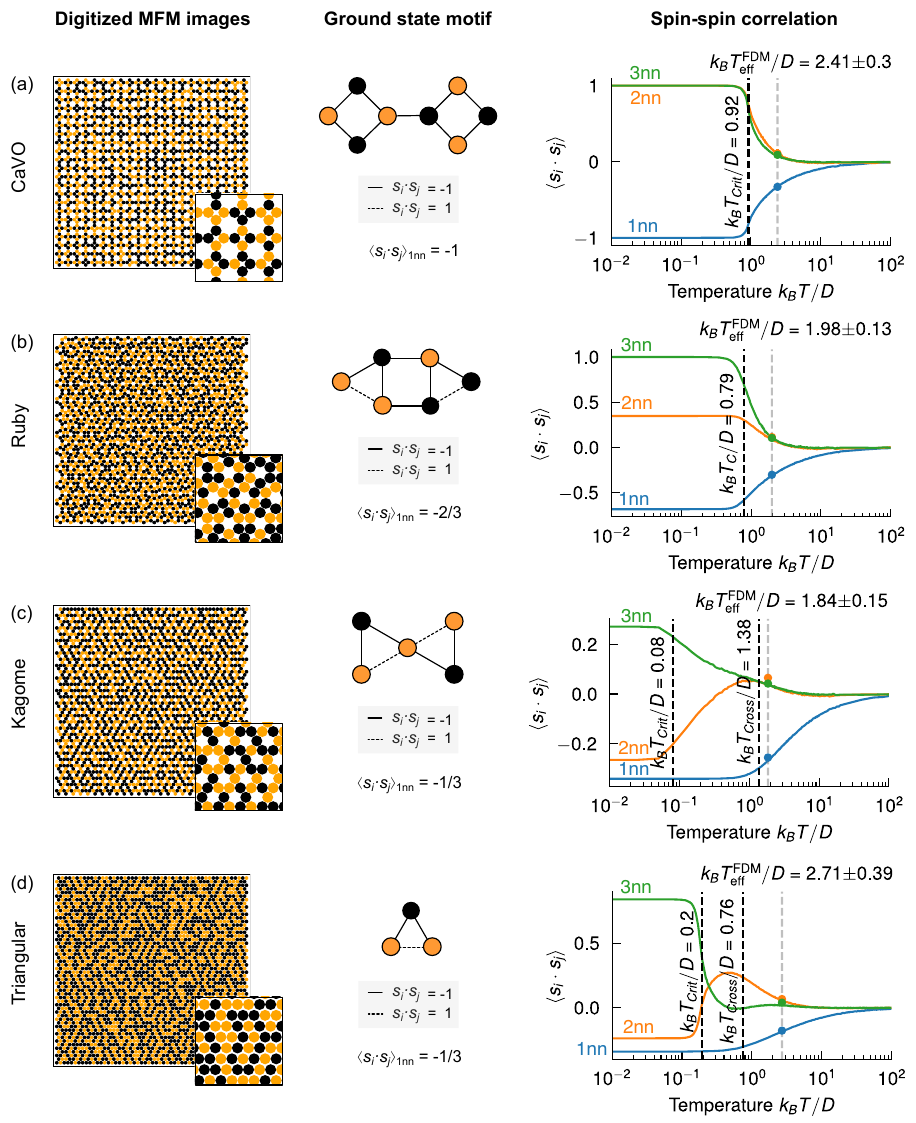}
    \caption{\label{fig:mfm}\textbf{Ordering in  four representative artificial spin lattices following demagnetization}. 
    Shown in the left hand column is the magnetic configuration in each lattice following demagnetization for (a) the bipartite CaVO lattice, (b) the frustrated Ruby lattice, and (c) and (d) the frustrated Kagome and Triangular lattices with two-step ordering. Here the MFM images were digitized and colored with orange (black) representing the magnetic moment pointing out of the plane (into the plane). 
    In the central column, we display the ground state motif that is expected at low temperature for each lattice with the corresponding value of $\langle \textbf{s}_i \cdot \textbf{s}_j \rangle$ given below each plaquette.
    For the unfrustrated CaVO lattice, the antiparallel alignment for all first-nearest-neighbor spins are energetically favorable, reflected by $\langle \textbf{s}_i \cdot \textbf{s}_j \rangle = -1$.
    For the frustrated Ruby lattice, the ground state motif consists of $6$ antiferromagnetically aligned pairs and $2$ ferromagnetically aligned pairs with a higher $\langle \textbf{s}_i \cdot \textbf{s}_j \rangle = -2/3$. 
    For the two-step ordered Kagome and Triangular lattice, the high frustration caused by unsatisfied interactions leads to $\langle \textbf{s}_i \cdot \textbf{s}_j \rangle = -1/3$.
    Shown in the right hand column are the spin-spin correlation curves for up to three nearest neighbors based on the Monte Carlo simulation results for the full-dipolar model. 
    The black dashed lines indicate the critical temperature $k_BT_{C}/D$, and the average value of the effective temperature $k_BT_{\mathrm{eff}}/D$ following demagnetization, based on the $15$ individual MFM measurements for each lattice type.
    The location of the critical or crossover temperatures is determined in each case from the temperature at which peaks occur in the specific heat capacity in Fig.~\ref{fig:PhaseDiagram}.
    The circular markers on the lines represent the calculated value of the average experimental spin-spin correlation with the error bars representing the standard deviation.
    The definitions of the nearest-neighbor spin pairs, up to the third-nearest-neighbor, are given for each of the Archimedean lattices in the Supplemental Material, Section~1~\cite{Supplementary}.}

\end{figure*}

We now focus on the magnetic configurations following demagnetization for four representative lattices: the bipartite CaVO lattice, the frustrated Ruby lattice, and both two-step frustrated lattices; the Triangular and Kagome lattices.
Neither the CaVO lattice nor the Ruby lattice have been experimentally investigated in an artificial spin system context.
The full results for all $11$~lattices are given in the Supplemental Material, Section~4~\cite{Supplementary}. 
For these four representative lattices, we display in Fig.~\ref{fig:mfm} digitized MFM images of the nanomagnet arrays obtained after applying a demagnetization protocol (left column), low-energy configurations for a plaquette (middle column), and the spin-spin correlations $\langle \textbf{s}_i\cdot\textbf{s}_j\rangle$ given by the full dipolar model (right column). 
The process by which the spin configuration is extracted from the digitized MFM images is described in Appendix~C.
The original MFM images for all lattices following the magnetic field protocol can be found in the Supplemental Material, Section~4~\cite{Supplementary}.
For the spin-spin correlations, the blue, orange and green curves are the first-, second- and third-nearest-neighbor spin-spin correlations as a function of temperature extracted from the full-dipolar-model Monte Carlo simulations. 
The average spin-spin correlations extracted from the MFM images are indicated in each case with circular markers with the corresponding colors. 
For each lattice, the black dashed vertical lines indicate the location of the critical or crossover temperatures $k_BT_{C}/D$, while the grey dashed vertical lines indicate the average effective temperature in the full-dipolar model, $k_BT_{\mathrm{eff}}^{\mathrm{FDM}}/D$, that each artificial spin system reaches following demagnetization.
As described in Appendix D, the effective temperature is defined as the temperature for which the mean square difference between the three experimental correlation values and the correlation values extracted from Monte Carlo simulations are minimized, and is a standard measure of the degree to which artificial spin systems reach their low-energy states.

The critical temperature is defined as the temperature below which the system transitions to a nonzero order parameter.
It corresponds to the temperature at which we observe a sharp peak in the heat capacity curve, which indicates a phase transition.
Equivalently, at a phase transition, there is some breaking of a global symmetry, which might be associated with the appearance of long-range order in the magnetic configuration.
As emphasized in Section~\ref{sec:phase_diagrams}, it remains uncertain whether the peak in each heat capacity corresponds to a true phase transition or a crossover for every lattice, both in the full-dipolar model and the first-nearest-neighbor model.
For the four representative lattices shown in Fig.~\ref{fig:mfm}, we make the following comments.
For the CaVO lattice, which belongs to the bipartite group, Monte Carlo simulations suggest the emergence of long-range order at the transition temperature, indicating this is likely a critical temperature. 
For the Ruby lattice, which belongs to the frustrated category, the ground state does not exhibit long-range order and so we are not able to definitively characterize the behavior in Fig.~\ref{fig:PhaseDiagram}(h) as either a phase transition or crossover. 
For the two-step frustrated lattices in the full-dipolar model, Smerald \emph{et al.}~\cite{2018Smerald} established that the high-temperature peak in the Triangular lattice  is a crossover, while the low-temperature peak marks a first-order phase transition. 
Similarly, Chioar \emph{et al.}~\cite{2016Chioar} concluded for the Kagome lattice that the high-temperature peak corresponds to a crossover, likely followed by a low-temperature phase transition.
To reflect these distinctions, we have labeled the black dashed vertical lines in Fig.~\ref{fig:mfm} as either  $T_{Crit}$  for critical temperatures or  $T_{Cross}$  for crossover temperatures, where we can be sure.

Looking at the graphs of spin-spin correlation versus temperature in Fig.~\ref{fig:mfm}, it becomes immediately apparent that none of the lattices displayed reach an effective temperature lower than $k_BT/D \approx 1$.
The implication of this is two-fold: first, we mainly work in a regime where the first-nearest-neighbor interaction is the dominant interaction.
This is because all the effective temperatures are on the scale of the first-nearest-neighbor interaction, $J_{\mathrm{1nn}} = 1$.
Second, we never encounter the phase transition in any of the lattices because this always occurs at $k_B T /D < 1$. 
Interestingly, for the Triangular lattice, the configurations obtained after demagnetization have a higher effective temperature, $k_BT_{\mathrm{eff}}^{\mathrm{FDM}}/D \approx 2.71$, as opposed to $k_BT_{\mathrm{eff}}^{\mathrm{FDM}}/D < 2.41$ for the other lattices.
On inspection of the digitized MFM images in Fig~\ref{fig:mfm}, we do not appear to achieve system-spanning order for any of the lattices.
At least, it is not possible to pick out a repeating (or translatable) pattern over the entire measured region, though some crystallites of magnetic order are present in some of the lattices. 

We now make a brief comment on each lattice in turn. 
For the bipartite CaVO lattice [Fig.~\ref{fig:mfm}(a)], there are small patches of the predicted antiferromagnetic ground state, which is built by tiling the motif shown in the middle column of Fig.~\ref{fig:mfm}.
The experimental spin-spin correlations agree extraordinarily well with those predicted by the Monte Carlo simulation at its effective temperature. 
Since the lattice admits a perfect antiferromagnetic ground state, the first-nearest-neighbor spin-spin correlation reaches $-1$ shortly after the critical temperature is traversed.
For the frustrated Ruby lattice [Fig.~\ref{fig:mfm}(b)], we also observe small patches of antiferromagnetic ordering. 
Additionally, the agreement between the experimental and simulated spin-spin correlations at the effective temperature is also excellent with minimal differences between the two. 
However, as a result of the frustrated nature of the artificial Ruby lattice, the first-nearest-neighbor spin-spin correlation is suppressed and the low temperature value that is reached is $-2/3$ as opposed to the value of $-1$ for the CaVO lattice.
This comes about because of the unsatisfied interactions within the triangular plaquettes that cannot be fully mitigated through the presence of the square plaquettes in the Ruby lattice.

For the two-step frustrated lattices, the situation is more nuanced. 
For the Kagome lattice [Fig.~\ref{fig:mfm}(c)], the tiling is such that the first-nearest-neighbor spin-spin correlation only reaches $-1/3$ at low temperatures indicating that the interactions on a plaquette cannot be fully satisfied and the system will never be fully ordered. 
The experimental spin-spin correlations match well with those predicted by Monte Carlo simulations and the effective temperature reached is $k_BT_{\mathrm{eff}}^{\mathrm{FDM}}/D=1.84$, which is the lowest $T_{\mathrm{eff}}$ out of the four example lattices.
However, the critical temperature of the Kagome lattice with the full-dipolar model, which we take to be the position of the low-temperature peak, is also the lowest $k_B T_C/D=0.08$.
This means that the Kagome lattice is much further from its phase transition, $k_B(T_{\text{eff}} - T_C) = 1.76$ than either the CaVO or Ruby lattices, where $k_B(T_{\text{eff}} - T_C)/D$ is equal to $1.475$ or $1.2$, respectively

For the Triangular lattice [Fig.~\ref{fig:mfm}(d)], the error bars in the spin-spin correlations are small but, in contrast to the Kagome lattice, the nanomagnet arrays appear to freeze out at a slightly higher effective temperature,  $k_BT_{\mathrm{eff}}^{\mathrm{FDM}}/D \approx 2.71$.
At this stage, it is important to note that all of the nanomagnet arrays based on the Triangular lattice were affected by a small dilation along the vertical direction, as seen in Fig.~\ref{fig:SEM}, which was introduced during the electron beam lithography.
This dilation affected only the Triangular nanomagnet arrays and the sizes of the nanomagnets themselves were unchanged.
A detailed analysis of its impact on the phase diagram and effective temperatures is provided in the Supplemental Material, Section 7. 
Here, we note that the effective temperatures obtained of such a dilated lattice remain within the standard deviation of those obtained for an ideal lattice. 
For the rest of this paper, we assume a perfect Triangular lattice.

In the Supplemental Material, Section~3~\cite{Supplementary}, we present the spin-spin correlations as a function of temperature up to the fifth-nearest neighbor for all $11$ Archimedean lattices, which are taken from the Monte Carlo simulations, using both the full-dipolar model and the first-nearest-neighbor model.
For both bipartite and frustrated lattices, the two models exhibit little qualitative difference in the behavior of the spin-spin correlations, reinforcing our earlier point that the phase diagrams for the lattices in each of these groups are remarkably similar---except that the critical temperatures given by the full-dipolar model are consistently lower than those given by the first-nearest-neighbor model.
However, the nearest-neighbor spin-spin correlations behave differently between the two models for the two-step frustrated lattices. 
Specifically, for both the Triangular and Kagome lattices, $\langle \mathbf{s}_i \cdot \mathbf{s}_j \rangle_{\mathrm{1nn}}$ reaches a lower value in the full-dipolar model. 
This reduced frustration is commensurate with the fact that a striped phase emerges in the full-dipolar model, which introduces a degree of long-range order.

\begin{figure*}
    \centering
    \includegraphics[width=\textwidth]{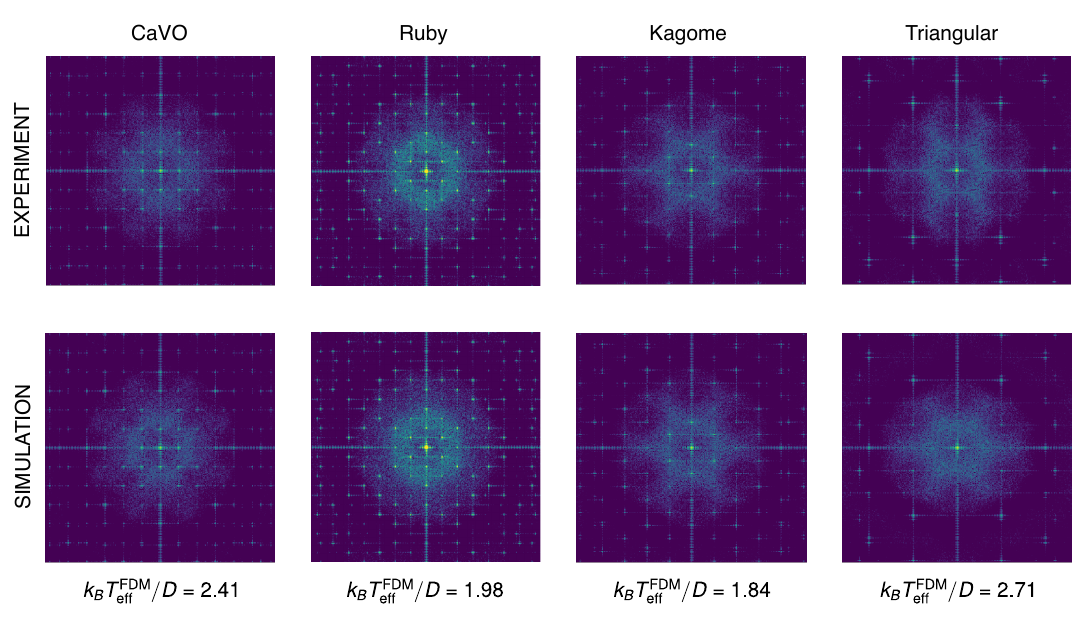}
    \caption{\textbf{Comparison of experimental and simulated magnetic structure factors for the four representative artificial spin lattices.} \textbf{Top row}: Magnetic structure factors determined from the experimental magnetic configurations following demagnetization. \textbf{Bottom row}: Corresponding magnetic structure factor calculated from the simulated configurations at the assigned effective temperature, as obtained using the full-dipolar model. The effective temperatures $k_B T_{\mathrm{eff}}^{\mathrm{FDM}}/D$ are indicated for each lattice. These are obtained by fitting the experimental data to the spin-spin correlations (see right-hand column of Fig.~\ref{fig:mfm}).}
    \label{fig:msf}
\end{figure*}

In Fig.~\ref{fig:msf}, we display the magnetic structure factors for the four representative lattices, determined from the experimental images (top row) and from a spin configuration from the Monte Carlo simulations incorporating the full-dipolar model at the calculated effective temperature (bottom row).
Further details about the calculation of the magnetic structure factor are given in Appendix~E.
A good agreement is seen for all lattices except the Triangular lattice, for which the structure factor obtained with Monte Carlo simulations appears slightly elongated along the horizontal axis while the magnetic structure factor of a typical experimental configuration appears elongated along the vertical axis. 

It should be noted that the magnetic configurations that were used as the basis for the experimental magnetic structure factor are taken from a single measurement, which is not precisely at the average effective temperature.
This experimental structure is then compared with a single magnetic configuration from Monte Carlo simulation taken exactly at $k_B T_{\mathrm{eff}}^{\mathrm{FDM}}$.
However, the true effective temperature of the experimental configuration still falls within one standard deviation of the average value. 
This can explain any small differences between the magnetic structure factors determined from MFM images and from Monte Carlo simulations.

\section{\label{sec:further_neighbor_interactions}Importance of including further neighbor interactions}
\begin{figure}
    \centering
    \includegraphics[width=85mm]{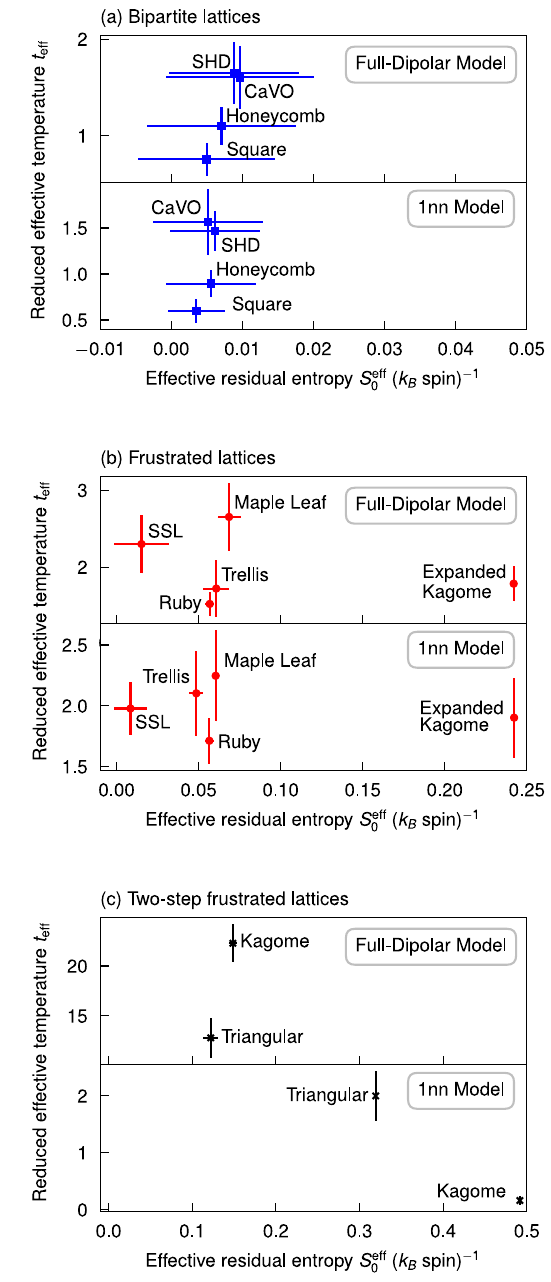}
    \caption{\label{fig:reduced_temp} \textbf{Reduced temperature as a function of effective residual entropy for the three groups of lattices.} The residual entropy was obtained using Monte Carlo simulations for the full-dipolar model and the first-nearest-neighbor model. The reduced effective temperature was calculated using the effective temperature taken from our experimental data and the critical temperature obtained from the simulated heat capacity curves. The three groups of lattices are: (a) bipartite lattices (blue), (b) frustrated lattices (red), and (c) the two-step frustrated lattices (black).}
\end{figure}

We now seek to understand the role of long-range interactions in determining the ordering in the artificial spin systems on Archimedean lattices. 
Our approach is based on relating the effective temperature we obtain in the experimental systems after demagnetization to the effective residual entropy predicted by our Monte Carlo simulations.
Since the phase transition occurs at a different temperature in each lattice, we define a reduced effective temperature $t_{\mathrm{eff}} = (T_{\mathrm{eff}} - T_C)/T_C$, where $T_{\mathrm{eff}}$ is the effective temperature and $T_C$ is the critical temperature at which the phase transition occurs.
In effect then, the reduced effective temperature is a measure of how close each lattice is to its respective critical point.
The smaller the value of $|t_{\mathrm{eff}}|$, the closer we are to the transition temperature.
The effective residual entropy $S^{\mathrm{eff}}_0$, which is a measure of the number of degenerate ground states, is obtained through an appropriate integration of the area under the curves of specific heat capacity versus temperature in Fig.~\ref{fig:PhaseDiagram} (see Appendix~B).
We describe this as an \emph{effective} quantity because it is not the true residual entropy. 
Rather, because we only use single spin flips in our Monte Carlo simulations, we expect the system to drop out of equilibrium at one point.
A more nuanced approach would employ some appropriate form of cluster~\cite{1989Wolff} or worm~\cite{2018Smerald} update in the Monte Carlo simulations that is tailored to each lattice in order to maintain ergodicity.
This would allow us to sample more precisely configurations at low temperatures.

In order to combine our experimental and simulation results, in Fig.~\ref{fig:reduced_temp} we plot the reduced effective temperature $t_{\mathrm{eff}}$, against effective residual entropy $S^{\mathrm{eff}}_0$, for the $11$~Archimedean lattices, comparing the results for the full-dipolar and first-nearest-neighbor models. 
We calculate the reduced effective temperature according to both the full-dipolar model and the first-nearest-neighbor model. 

The bipartite lattices [Fig.~\ref{fig:PhaseDiagram}(a)] all have an $S^{\mathrm{eff}}_0$ close to zero as a result of the finite (two-fold) degeneracy of their ground states, and subsequently have the lowest $t_{\mathrm{eff}}$ of any of the three groups of lattices. 
The numerical values for the effective temperatures and effective entropies for both models for all the $11$~Archimedean lattices are given in the table in the Supplemental Material, Section~6~\cite{Supplementary}.
In general though, demagnetization in a magnetic field still proves insufficient to bring these systems close to their phase transition with $t_{\mathrm{eff}}\textgreater 0.7$ and, as we remarked earlier, only patches of antiferromagnetically-ordered regions are observed.
For this group of lattices, there is little difference between the results for the full-dipolar model and the first-nearest-neighbor model.

In Fig.~\ref{fig:reduced_temp}(b), we show a plot of $t_{\mathrm{eff}}$ versus $S_{0}^{\mathrm{eff}}$ for the five frustrated lattices. 
On average, these lattices have a higher $t_{\mathrm{eff}}$ than that of the bipartite lattices, as well as a non-zero $S^{\mathrm{eff}}_0$. 
For these lattices, not all of the interactions can be satisfied simultaneously, and thus there are many low energy states that lead to the emergence of non-zero $S_{0}^{\mathrm{eff}}$. 
For most of the lattices in this group $S_{0}^{\mathrm{eff}}$ is below $0.1 k_B$~per spin with the exception of expanded Kagome lattice for which $S^{\mathrm{eff}}_0\approx0.24 k_B$~per spin.
In addition to the increase in $t_{\mathrm{eff}}$ when going from the bipartite to frustrated lattices, we also observe a slight increase in the spread of $t_{\mathrm{eff}}$
The tendency towards larger $t_{\mathrm{eff}}$ is due to the larger proportion of unsatisfied interactions present in the frustrated lattices leading to a high degeneracy of low energy states so that reaching the ground state is more challenging than for the bipartite group.
However, while frustration plays an important role in these lattices, we are still able to reach effective temperatures with the same order of magnitude as for the bipartite lattices irrespective of whether these are calculated according to the full-dipolar model or first-nearest-neighbor model. 
As we showed in Fig.~\ref{fig:PhaseDiagram} for the frustrated lattice, the heat capacity versus temperature curves for the full-dipolar model and the first-nearest-neighbor model are very similar with only a single peak that occurs at similar temperatures.
It therefore appears that the inclusion of further-neighbors has little effect on the overall ordering of these lattices.
As further evidence of this, we see in the Supplemental Material, Section~5~\cite{Supplementary} that the magnetic structure factors for the two models at $T=0$ look very similar.

Finally, for the two-step frustrated lattices [Fig.~\ref{fig:reduced_temp}(c)], the Triangular and Kagome lattices, we observe a significant effect on including interactions beyond the first-nearest-neighbor interactions.
For these two lattices, we take $T_C$ as the temperature at which the lower of the two peaks in the heat capacity occurs.
The rationale behind this is that the upper peak is a crossover rather than a true phase transition, because no global symmetry is broken.
Specifically, for the Triangular and Kagome lattices, the upper peak corresponds to the emergence of a rule---two-spins-up/one-spin-down or vice versa---governing the allowed low-energy configurations within a triangular plaquette, making the all-up or all-down states less likely. 
In addition, the broadness of the upper peak in both lattices suggests a crossover because it reflects a wide temperature range over which the system transitions from the high-temperature regime, where all plaquette states are equally probable, to a lower-temperature regime where only specific plaquette rule-obeying states are energetically favored.
The exact nature of the low-temperature phase transition in the Kagome lattice has not been established, though there exists a promising candidate for its ground state based on an interlocking seven-shaped spin motif~\cite{2016Chioar}.

For the Triangular lattice, Smerald \emph{et al.}~\cite{2018Smerald} established that the lower peak corresponds to a first-order phase transition from a spin liquid phase above the transition to a striped phase below.
The presence of the first-order transition peak at such low temperature explains why we obtain a large $t_{\mathrm{eff}}$ in the demagnetized systems, which is much higher than in the other nine lattices. 
For the Triangular lattice, the reduced effective temperature is $t_{\mathrm{eff}} = 12.8$ while for the Kagome lattice it is $t_{\mathrm{eff}} = 22.8$.
Similarly, the effective residual entropy for these lattices, $S_{0}^{\mathrm{eff}} = 0.09 k_B$ per spin for Triangular lattice and $S_{0}^{\mathrm{eff}} = 0.15 k_B$ per spin for Kagome lattice, is higher than those in the frustrated group, which all have $S_{0}^{\mathrm{eff}} \textless 0.07 k_B$ per spin (with the exception of the Expanded Kagome lattice for which $S_{0}^{\mathrm{eff}} = 0.24 k_B$ per spin).
It is apparent that reaching low energy state in the two-step lattices poses a significant challenge, resulting in large effective temperatures.
For the two-step frustrated lattices, the experimental nanomagnet arrays appear to freeze out in the vicinity of the crossover temperature associated with the broad peak appearing at higher temperatures in Fig.~\ref{fig:PhaseDiagram}.
This means that observing an ordered state for any of the two-step frustrated lattices may require an unconventional approach that is beyond the scope of this work.
Possible approaches for the experimental spin systems with in-plane magnetized nanomagnets have been previously demonstrated, such as modifying the individual elements making up the lattices \cite{2022Hofhuis,2022Yue}. 

Lastly, an important observation can be made concerning the effective residual entropy in the frustrated systems and the two-step frustrated systems obtained using the two different models.
For the two-step frustrated lattices, we observe a decrease of effective residual entropy when the first-nearest-neighbor model is extended to include long-range interactions.
This in turn suggests that the full-dipolar model gives a more ordered ground state. 
This is in contrast to the frustrated lattices for which the effective residual entropy increases when the long-range interactions are accounted for, as shown in Table I in the Supplemental Material, Section~6~\cite{Supplementary}.
This would suggest that these lattices become disordered for the full-dipolar model.
However, it is more likely that reaching a true ground state becomes more difficult due to the additional long-range interactions.
The higher level of frustration brought about by the additional long-range interactions prevents the system from reaching a fully ordered state, thus retaining a higher effective residual entropy when only single spin flips are used.

\section{\label{sec:conclusions}Conclusions}
We have engineered artificial spin systems based on the $11$~Archimedean lattices comprising arrays of Co/Pt circular nanomagnets with perpendicular magnetic anisotropy. 
The nanomagnet arrays were prepared in low-energy states with a demagnetization protocol using an alternating magnetic field. 
From images of the exact microstate of each lattice with MFM, we calculated the experimental spin-spin correlations and compared these to the correlations predicted by using Monte Carlo simulated anneals.
This enabled us to assign effective temperatures to each lattice, demonstrating that, irrespective of the lattice, the system freezes just above its critical or crossover temperature, and is unable to fully reach these transitions.

Indeed, it is difficult to obtain configurations with effective temperatures $k_B T /D < 1$.
This temperature threshold, which is similar across all $11$~lattices,  is likely to correspond to the energy barrier to macrospin reorientation in a single nanomagnet, rather than, say, the presence of quenched disorder in the samples.
This hypothesis is supported by the analysis we present in the Supplemental Material, Section~3~\cite{Supplementary}, where we perform Monte Carlo simulations for the Square, Trellis, Triangular and Kagome lattices with varying levels of quenched disorder, introducing a Gaussian distribution of intermagnet interactions around their nominal value.
The amount of disorder in a given Monte Carlo simulation is then reflected by the standard deviation of the Gaussian.
We find that, across the lattices, the high-temperature portion of the phase diagram, $k_B T /D >1$, appears relatively immune to the presence of even significant amounts of disorder, e.g. up to $20\%$. 
Instead, we find that the presence of disorder affects only the low-temperature peaks in the Triangular and Kagome lattices, causing it to disappear (broaden) in the Triangular (Kagome) lattice.

We were able to sort the lattices into three groups:
\begin{enumerate}
    \item bipartite lattices that host a perfectly antiferromagnetic ground state;
    \item frustrated lattices, whose phase diagram features a single transition or crossover; and
    \item two-step frustrated lattices, where the specific heat capacity has two peaks.
\end{enumerate}
Perhaps as expected, the bipartite lattices displayed the lowest reduced effective temperatures, with $t_{\mathrm{eff}}$ of the frustrated lattices being slightly higher.
The two-step frustrated lattices had the highest $t_{\mathrm{eff}}$, confirming the difficulty in accessing the low-energy configurations in such highly frustrated systems.
We attempted to quantify the frustration in these systems through their effective residual entropy. 
Here we distinguish between this \textit{effective} residual entropy, which is determined from our Monte Carlo simulations based on single spin flips only, and the \textit{true} residual entropy, which requires a more nuanced way to sample the possible configurations in highly frustrated systems (see Appendix~B for more information).

Furthermore, we established the importance of including further-neighbors interactions when calculating the phase diagram of certain Archimedean lattices, distinguishing between two cases: the first-nearest-neighbor model and the full-dipolar model, in which every spin is coupled to all of the others. 
For most of the lattices, the introduction of further neighbor interactions does not affect the phase diagram greatly in the temperature range in which our experimental samples freeze out.
For the two-step frustrated lattices, the Triangular and Kagome lattices, however, there is a difference between the first-nearest-neighbor model and the full-dipolar model, in that the full-dipolar model features a low-temperature phase transition, which is likely to be first-order in both cases. 
The existence of this extra low-temperature peak when all dipolar interactions are accounted for was previously noted in References~\cite{2016Chioar, 2018Smerald}. 
Here, our results show that this difference in thermal behaviour leads to a higher reduced effective temperature, but a lower effective residual entropy.

In summary, we provide the most complete classification of artificial spin systems with perpendicular magnets arranged on the Archimedean lattices.
We have shown how long-range interactions influence the ordering in systems with varying levels of frustration, going from the bipartite to the highly-frustrated lattices. 
Reaching lower effective temperatures will require precise tuning of the material parameters to control the energy barrier required for switching and the intermagnet coupling~\cite{2024Kurenkov}.
In addition to achieving lower energy states, this will provide a way to observe  thermal relaxation at room temperature, which is something already achieved for in-plane systems~\cite{2013Farhan, 2017Gliga, 2020Hofhuis, 2022Hofhuis}.

Many of these Archimedean artificial spin systems have been realized here for the first time.
Future observations of their thermodynamics, in particular how they order through a crossover or first-order phase transition, will provide ample scope for future research.
This may be especially true for those Archimedean lattices that feature both even- and odd-sided polygons, which results in a mixture of frustrated and unfrustrated motifs.
However, the Kagome and Triangular lattices are likely to continue to fascinate, as the precise nature of their low-temperature states remains a topic of ongoing research, with some promising candidates identified~\cite{2016Chioar}.

Finally, the insights gained into the delicate interplay between lattice geometry and long-range interactions are critical for advancing these systems toward applications in novel computing schemes.
For example, the computational capacity of spin lattices can be harnessed by tailoring the lattice geometry, so that they respond to particular inputs, such as field cycles, in specific ways~\cite{2021Paterson, jack, 2024Jensen}.
This would provide a route to the the low-power processing of data envisaged by reservoir computing.

The dataset supporting this study is available in an online repository~\cite{Zenodo}.

\begin{acknowledgments}
We thank Anja Weber and  Vitaliy Guzenko for their help with sample fabrication, and Thomas Jung for access to the MFM tool. 
We would also like to thank Afonso Dos Santos Rufino at EPFL and Jeanne Colbois at National University of Singapore for helpful discussions. 
We are grateful for financial support for this work from the Swiss National Science Foundation (project no. 200020\_200332). 
We would also like to acknowledge the support from the EU FET-Open RIA project SpinENGINE (Grant no. 861618).
\end{acknowledgments}

\appendix
\section{Fabrication}

Artificial spin systems with circular nanomagnets arranged on the $11$~Archimedean lattices were fabricated with electron beam lithography on the same chip.
This minimized potential sources of variation due to the fabrication process and/or the measurement conditions.
The arrays of nanomagnets were small enough  (\textless $10\times10~\upmu$m$^2$) to fit within a single scanned image of the MFM, allowing us to capture  the magnetic configuration of all of the nanomagnets within a single image. 
They were fabricated with electron beam lithography from a Co/Pt multilayer deposited on a silicon oxide substrate with UHV sputtering, with the exact film composition being $\mathrm{Ta(4)/Pt(3)/[Co(1.1)/Pt(0.2)]_{2}/Co(1.1)/Ru(2)}$.
The numbers in brackets correspond to the thickness of the layers in nm.

Using a superconducting quantum interference vibrating sample magnetometer (SQUID-VSM), we have measured the bulk magnetic properties of the film for magnetic fields applied parallel and perpendicular to the sample plane, confirming the PMA nature of the stack. 
The hysteresis curves from the SQUID-VSM measurements can be found in the Supplemental Material, Section~2~\cite{Supplementary}.

The nanomagnet diameter was chosen to be $200$~nm to ensure that the nanomagnets are single domain, with a lattice constant of $225$~nm, giving an intermagnet separation of $25$~nm.
The smaller the separation, the larger the dipolar coupling between the nanomagnets.
This is the smallest nanomagnet separation for which we can be sure that adjacent nanomagnets are completely separated. 

In more detail, the nanomagnet arrays were fabricated by first depositing the magnetic multilayer onto a silicon substrate using DC magnetron sputtering at a base pressure of $2.5\times10^{-8}$~mbar. 
The film was pre-baked at $125\degree$C for $5$~minutes to help with the adhesion between the top surface of the film and the electron-beam resist.
Once cooled down, the film was spin coated with hydrogen silsesquioxane (HSQ) electron-beam resist (XR-1541~2$\%$) at $4000$~rpm for $60$~s to obtain a total resist thickness of ~$\sim50$~nm. 
This was followed by electron beam exposure of the HSQ resist with the desired patterns of arrays of dots using a Raith EBPG$5000$PlusES electron beam writer. 
Due to the negative polarity of the resist, the developed resist was an array of HSQ dots that was subsequently used as a mask for milling the Co/Pt multilayer with Ar ions, leaving behind the arrays of circular nanomagnets. 
The ion milling took $23$~s, during which the sample was rotated at $20$~rpm and tilted at $15$\degree~to the Ar beam.
The Ar beam voltage was $400$~V; the flow of the gas was kept to $8$~sccm; and the ion current was $122$~mA.
Finally, the resist on top of the nanomagnet arrays was removed using a buffered oxide etch solution of hydrofluoric acid. 
The circular nanomagnets appeared with a uniform black or white contrast in the MFM images, indicating that they were single-domain and magnetized out-of-plane. 
Using scanning electron microscopy, we confirmed the diameter and separation of the nanomagnets for each lattice.

\section{Monte Carlo Simulations}

To capture the thermal behaviour of each of the Archimedean spin lattices, Monte Carlo simulations were performed using the Metropolis-Hastings algorithm with single spin flip dynamics. 
Each nanomagnet was approximated as an Ising point dipole (or spin) that can take one of the two out-of-plane directions.
The Hamiltonian of the system was

\begin{equation}
\label{eq:dipolar_hamiltonian}
\mathcal{H}=D\sum _{i\neq j}\dfrac{\textbf{s}_{i}\cdot \textbf{s}_{j}}{r_{ij}^{3}},
\end{equation}

\noindent which is the simplified form of the dipolar interaction appropriate for out-of-plane spins. 
In Eq.~(\ref{eq:dipolar_hamiltonian}), the spin \textit{i} at position ${\textbf{r}_{i}}$ has magnetic moment ${\textbf{s}_{i}}$. 
The connecting vector $\textbf{r}_{ij}=\textbf{r}_{j}-\textbf{r}_{i}$ points from spin \textit{i} to spin \textit{j}. 
The energy scale of the interactions is set through the dipolar constant, $D=\mu _{0}\left( M_{S}V\right) ^{2}/a^{3}$, where $M_{S}$ and $V$ are the saturation magnetization and volume of each nanomagnet, respectively. 
We work in reduced units throughout this paper so that the first-nearest-neighbor interaction strength,  $J_{1nn} = D = 1$. 
The temperature is then quoted in dimensionless units, i.e., $k_B T / D$.

In this work, we have also considered the effect of truncating the dipolar interaction to the first-nearest-neighbors.
This cutoff allows us to map between a first-nearest-neighbor antiferromagnetic Ising-like model and the full-dipolar model.
Such a cutoff has been shown to affect the critical behaviour of system, especially at low temperatures~\cite{2014Chioar}. 

For a system of $N$ spins, a single Monte Carlo step involves \textit{N} attempted single spin flips. 
At each temperature, $10^4$ Monte Carlo steps are used to equilibrate the system before $10^4$ steps are used for averaging. 
The configuration from the previous temperature is then used as the starting state for the next temperature point. 
The heat capacity is determined from the first and second moments of the energy using 
\begin{equation}
    c_{V}=\frac{1}{N}\dfrac{\langle E^{2}\rangle -\langle E\rangle ^{2}}{k_{B}T^{2}},
\end{equation}
which has units of ($k_B$ spin$^{-1}$).

To obtain a measure of the level of frustration in the lattices, we define an \emph{effective} residual entropy $S^{\mathrm{eff}}_0$, which is the residual entropy of the state reached when only single spin flips are used in the Monte Carlo simulations.
This gives a measure of the degeneracy of states at zero temperature. 
The quantity $S^{\mathrm{eff}}_0$ is obtained by integrating the heat capacity $C_V$ and subtracting this value from the entropy of a spin in the high-temperature limit $S = k_B \ln(2)$ as follows
\begin{equation}
    S^{\mathrm{eff}}_{0} = k_B \mathrm{ln(2)} - \int_{0}^{100} \dfrac{C_V}{T}\,\mathrm{d}T. 
\end{equation}
\noindent where the temperature limits for the integral correspond to the final and starting temperatures in our Monte Carlo simulated anneal. 
The greater the value of $S^{\mathrm{eff}}_0$, the higher the degeneracy of the state reached and thus the level of frustration.\\

The data points from the Monte Carlo simulations reflect the average of at least $20$ independent runs. 
Error bars are taken as the standard deviation of this average. 
The temperature is first decreased linearly from $k_BT/D = 100$ to $k_BT/D = 1$, before decreasing it logarithmically, with a final temperature at $k_BT/D = 0$.
For all lattices, the specific heat capacity evaluated at the highest and lowest temperatures is very small, with any transitions happening somewhere within the temperature range $0 < k_B T /D < 100$. 
This ensures that our estimate of the effective residual entropy 
is more accurate as it depends on integrating over a wide temperature range of the specific heat curve, including the peaks associated with the phase transitions and crossovers.
In principle, a small correction arises from the behavior of the specific heat between the highest simulated temperature and infinity. However, by selecting the highest annealing temperature $k_B T /D =100$, where the specific heat is already small, any such correction becomes negligible.
At least 100 data points are distributed in each temperature decade. 
In total, $254$ temperature points are used.
Periodic boundary conditions were implemented by creating periodic copies of the system out to $15$ times the system size.

For those lattices that are highly frustrated, e.g. the Triangular or Kagome lattice, it is likely that single spin flips are not sufficient to completely equilibrate the system. 
This problem will be especially acute in the vicinity of the critical temperature, where the autocorrelation time is large.
The autocorrelation time in the context of artificial spin systems is a measure of the timescale over which the spin configuration of a system remains correlated with itself.
A short autocorrelation time indicates that the system evolves rapidly and configurations become decorrelated quickly, whereas a long autocorrelation time suggests slower dynamics and more persistent correlations, meaning that the estimates of our observables, e.g. heat capacity, will not be accurate.
As we mentioned in the main text, a more sophisticated approach would be to implement non-local updates, such as cluster or worm algorithms. 
However, as we have demonstrated, all experimental systems appear to freeze out at a temperature where single spin-flips still capture the essential physics, and so this does not affect our assignment of effective temperatures. 

\section{Extracting the Spin Configurations from MFM Measurements}
MFM measurements were performed using a Bruker Veeco Dimension 3100. 
For this, commercially available low-moment Bruker MESP-LM-V2 tips were used.
The scan size was \SI{15}{\micro\metre}$\times$\SI{15}{\micro\metre}, with a typical lift height of \SI{40}{\nano\metre} and a scan frequency of \SI{0.5}{\hertz}.
There were either 256 or 512 pixels per line, and each image took approximately 20 minutes to acquire.
Since MFM is sensitive to the out-of-plane component of the stray field gradient, each single-domain circular nanomagnet appears as either a bright or a dark disc.
Bright (dark) contrast within a nanomagnet means that there is a repulsive (attractive) interaction between the magnetic moment of the nanomagnet and the magnetic moment of the tip.
Equivalently, regions of bright or dark contrast correspond to the macrospin moment and the tip moment being anti-parallel or parallel, respectively.  
Without \emph{a priori} knowledge of the exact direction of the magnetic moment of the tip, an absolute direction for every macrospin cannot be assigned.
Instead, the magnetic configuration of the nanomagnet array can be determined up to a global spin flip. 
This is possible because the magnetic moment of the tip does not change during the course of a scan.
We adopt the convention that bright contrast corresponds to an up macrospin, i.e. $s = + 1$.

Each MFM image was post-processed using a custom Python code.
First, a Savitzky-Golay filter was applied to remove the non-linear background from each row in both the topography and magnetic phase channels.
Then, the position of the lattice in each image was identified, which enabled classification of all pixels in the image as either part of a nanomagnet or the background.
A local threshold, based on the average intensity of nearby background pixels, was then applied to the nanomagnet pixels in the magnetic phase channel in order to digitize the image.
To automatically read in the orientation of each macrospin, our code calculated the average intensity of the pixels within each nanomagnet and, when compared to the overall mean, used this as a figure of merit to classify the nanomagnet as bright (intensity greater than the overall mean, corresponding to an up macrospin) or a dark (intensity less than the overall mean, corresponding to a down macrospin).

\section{Assigning the Effective Temperature}

For each Archimedean lattice, the effective temperature in the full-dipolar and first-nearest-neighbor models is determined by comparing the experimental spin-spin correlations with those obtained from the Monte Carlo simulations. 
More precisely, the effective temperature for a given model is defined as the temperature that minimizes the sum of the mean-squared deviations between the experimental correlations and the Monte Carlo predictions for the first three nearest-neighbor spin-spin correlations. 
The effective temperatures as a function of separation, computed for both models, are provided in Supplemental Material, Section 4.

\section{Magnetic Structure Factors}

The intensity of the magnetic structure factor $I(\mathbf{q})$ is calculated by obtaining the spin-spin correlations in real space, in our case via the MFM measurements, followed by the Fourier transform to map the spin correlations to reciprocal space.
For a system of spins $\mathbf{s}_{i}$ at lattice positions $\mathbf{r}_{i}$, the intensity of the magnetic structure factor is given by 

\begin{equation}
    I(\mathbf{q}) = \sum_{i,j} \textbf{s}_i \cdot \textbf{s}_j  e^{iq \cdot (r_{i}-r_{j})}
\end{equation}

\noindent where $\mathbf{q}$ is the wave vector at a point in reciprocal space, and $\mathbf{s}_i$ and $\mathbf{s}_j$ are the normalized moments of two spins at positions $r_i$ and $r_j$, respectively.
The double sum is taken over all pairs of spins in the system.

The magnetic structure factor is used to characterize the spatial distribution and correlations of magnetic moments.
For ordered states, the magnetic structure factor contains Bragg peaks that appear for specific wavevectors and represent dominant periodicities and correlations in the spin configuration.
The magnetic structure factor can therefore be used to indicate the presence of long-range correlations.
For disordered systems the magnetic structure factor becomes diffuse, and thus can be used to indicate a reduction of long-range ordering. 
This can help to determine which lattices have long-range or short-range magnetic order.
The magnetic structure factors determined from the MFM images and Monte Carlo simulations are discussed in more detail in the Supplemental Material, Section~5~\cite{Supplementary}.
Here we compare the experimental magnetic structure factor to that of a configuration taken from Monte Carlo simulations at $T_{\mathrm{eff}}$.
We also use the simulated magnetic structure factors to highlight the differences in magnetic ordering when using the full-dipolar model compared with the first-nearest-neighbor model. 
The differences are most apparent when comparing the magnetic structure factors for the ground state taken from Monte Carlo simulations of the Triangular and Kagome lattices.
In particular, for these two-step frustrated lattices, the peaks in the magnetic structure factor are sharper for the full-dipolar model than those for the first-nearest-neighbor model.

Finally, in Section 5 of the Supplemental Material~\cite{Supplementary}, we compare the real-space magnetic configurations and magnetic structure factors at $T = 0$, obtained from Monte Carlo simulations, with those at the experimentally-measured effective temperature. 
In general, the real-space configurations at $T = 0$ exhibit more long-range order than those at $T_{\mathrm{eff}}$.
This is most obviously seen across all $11$~lattices by the reduction in diffuse intensity in the ground state magnetic structure factors compared with those at the effective temperature. 

%\bibliography{References}% Produces the bibliography via BibTeX.

\providecommand{\noopsort}[1]{}\providecommand{\singleletter}[1]{#1}%

\end{document}